\documentclass{Interspeech}
\usepackage{multirow}

\interspeechcameraready

\title{StreamFlow: Streaming Flow Matching with Block-wise Guided Attention Mask for Speech Token Decoding}

\author[affiliation={1}]{Dake}{Guo}
\author[affiliation={1}]{Jixun}{Yao}
\author[affiliation={1}]{Linhan}{Ma}
\author[affiliation={1}]{He}{Wang}
\author[affiliation={1,*}]{Lei}{Xie}

\affiliation{}{Audio, Speech and Language Processing Group (ASLP@NPU), School of Computer Science, Northwestern Polytechnical University,}{Xi'an, China}
\email{guodake@mail.nwpu.edu.cn, lxie@nwpu.edu.cn\thanks{* Corresponding author.}}


\usepackage{comment}

\begin{document}

\maketitle

\begin{abstract}

Recent advancements in discrete token-based speech generation have highlighted the importance of token-to-waveform generation for audio quality, particularly in real-time interactions. Traditional frameworks integrating semantic tokens with flow matching (FM) struggle with streaming capabilities due to their reliance on a global receptive field. Additionally, directly implementing token-by-token streaming speech generation often results in degraded audio quality.
To address these challenges, we propose StreamFlow, a novel neural architecture that facilitates streaming flow matching with diffusion transformers (DiT). To mitigate the long-sequence extrapolation issues arising from lengthy historical dependencies, we design a local block-wise receptive field strategy. Specifically, the sequence is first segmented into blocks, and we introduce block-wise attention masks that enable the current block to receive information from the previous or subsequent block. These attention masks are combined hierarchically across different DiT-blocks to regulate the receptive field of DiTs.
Both subjective and objective experimental results demonstrate that our approach achieves performance comparable to non-streaming methods while surpassing other streaming methods in terms of speech quality, all the while effectively managing inference time during long-sequence generation. Furthermore, our method achieves a notable first-packet latency of only 180 ms.\footnote{Speech samples: https://dukguo.github.io/StreamFlow/}

\end{abstract}

\section{Introduction}

Recent advancements in speech generation and dialogue systems, such as GPT-4o~\cite{DBLP:journals/corr/abs-2410-21276} and Moshi~\cite{DBLP:journals/corr/abs-2410-00037}, have achieved remarkable multimodal speech interaction capabilities. These systems enable real-time conversations with humans, demonstrating near-human fluency. Such human-computer interaction requires systems that can perform reasoning directly with speech and generate output in real-time. Speech synthesis, a key technique in human-computer interaction systems, has received widespread attention, and the demand for real-time streaming speech generation has been growing.


Although conventional speech synthesis approaches~\cite{DBLP:conf/iclr/0006H0QZZL21,DBLP:conf/icml/KimKS21} can support streaming synthesis, the synthesis samples are not satisfactory in terms of voice quality, similarity, and prosody.
Inspired by the success of LLMs for text, the next-token prediction paradigm has been extended to tasks involving speech modalities, particularly in speech synthesis. Neural codec-based language models (Codec-LM)~\cite{DBLP:journals/corr/abs-2410-23815,DBLP:journals/corr/abs-2305-07243,DBLP:journals/corr/abs-2402-08093} use an autoregressive language model to predict discrete speech tokens extracted from the neural codec. With a large dataset of text-speech pairs, Codec-LM exhibits emergent in-context learning capabilities, enabling the model to perform transcript-conditioned speech continuation tasks with just a short speech prompt. Codec-LM not only integrates seamlessly into the LLM ecosystem but also enhances the interactive experience of speech agents, enabling real-time generation of high-quality speech content~\cite{DBLP:journals/corr/abs-2411-13577}. Therefore, efficiently and effectively streaming the decoding of speech tokens generated by Codec-LM has become a crucial research challenge in the field of speech synthesis.

One of typical Codec-LMs, such as CosyVoice~\cite{cosy} and GLM-4-Voice~\cite{DBLP:journals/corr/abs-2412-02612}, explore the use of automatic speech recognition (ASR) encoders to discretize speech tokens with richer semantic content. These tokens primarily contain semantic information with minimal acoustic details, making them more compatible with LLMs and enhancing generation efficiency. However, the lack of sufficient acoustic details in ASR tokens presents challenges when reconstructing the generated tokens into speech signals. To address this, more powerful generative models such as conditional flow matching (CFM)~\cite{DBLP:conf/iclr/LipmanCBNL23} are used to fill in the missing acoustic details, improving the accuracy of speech token decoding, which plays a crucial role in the Codec-LM-based framework. 

Although combining ASR tokens with CFM can balance speech synthesis naturalness and quality, the original CFM still struggles to meet real-time decoding requirements in streaming scenarios due to its reliance on the global receptive field, often causing significant generation delays. However, directly implementing streaming generation with token-by-token may lead to a decline in audio quality. Some studies~\cite{DBLP:journals/corr/abs-2412-02612} attempt to split speech token sequences into smaller chunks, achieving streaming speech generation through segmented generation. Although this approach improves generation speed and supports chunk-level streaming synthesis, it often leads to issues such as phase conflicts and popping sounds due to the lack of connection between chunks. To address this, CosyVoice2~\cite{cosy2} introduces causal block attention masks and causal convolutions to control the receptive field of flow matching, ensuring that each chunk only references historical information, enabling chunk-by-chunk generation. This method addresses issues to some extent, but as historical information increases during the generation of longer sequences, the inference time grows significantly, impacting real-time performance. In long conversation scenarios, generation efficiency could become a bottleneck.

To address the above issues and further enhance the efficiency and effectiveness of streaming generation, we propose StreamFlow, a streaming flow matching model with diffusion transformers (DiTs)~\cite{DBLP:conf/iccv/PeeblesX23} for streaming speech synthesis in the decoding phase of Codec-LM. We introduce a block-wise guided attention mask to effectively capture locality information across blocks, significantly improving the efficiency of long-sequence generation. Specifically, the sequence is first segmented into blocks, and we design three fundamental block-wise attentions to constrain the information captured during generation. These attentions are hierarchically combined across different DiT blocks to regulate the receptive field. The proposed block-wise guided attention mask prevents phase conflicts and popping sounds while enabling efficient and stable long-sequence streaming speech generation. Both subjective and objective experimental results demonstrate that our proposed approach achieves comparable performance to non-streaming methods and outperforms other streaming methods in speech
quality, with inference time effectively controlled during long-sequence generation. Furthermore, our approach achieves a notable first-packet latency of only 180ms.


\section{Preliminaries}
\subsection{Flow Matching}
In this section, we provide a brief description of flow matching (FM). The goal of FM is to match a probability path that transforms a data distribution \( p_t \) into a simple distribution \( p_0 \) (typically \( p_0 \sim \mathcal{N}(0,1) \)).
It is closely related to Continuous Normalizing Flows(CNFs)~\cite{DBLP:conf/nips/ChenRBD18} but is trained much more efficiently in a simulation-free fashion, similar to the typical training paradigm in diffusion probabilistic models (DPMs).

We can define the flow \( \phi: \mathbb{R}^d \to \mathbb{R}^d \) as the transformation that maps between two density functions, constraint by the following ODE:
$$
d\phi _t (x) = v_t(\phi _t(x))dt, t \in [0,1]; \phi_0(x)=x, 
$$
where \( v_t(\cdot) \) represents the time-dependent vector field,  which defines the generation path \( p_t \) as the marginal probability distribution of data points \(x\). Sampling can be performed from the approximate data distribution \( p_1 \) by solving the initial value problem in the equation.

We assume a vector field as  \( u_t \), which generates a probability path \( p_t \) from \( p_0 \) to \( p_1 \), the FM loss is defined as:  
$$
\mathcal{L}_{\text{FM}}(\theta) = \mathbb{E}_{t,p_t(x)} ||  u_t(x)-v_t(x;\theta) ||^2,
$$
where \( v_t(x; \theta) \) is a neural network parameterized by \( \theta \). However, this is challenging to implement in practice, as obtaining the vector field \( u_t \) and the target probability distribution \( p_t \) is nontrivial.
Therefore, Conditional Flow Matching (CFM) loss is designed in an alternative way as:
$$
\mathcal{L}_{\text{CFM}}(\theta) = \mathbb{E}_{t,q(x_1),p_t(x|x_1)}||u_t(x|x_1)-v_t(x;\theta)||^2.
$$
CFM replaces the intractable marginal probability density and vector field with conditional probability density and conditional vector field. The key advantage is that these conditional densities and vector fields are readily available and have closed-form solutions. Furthermore, it can be proven that the gradients of \( \mathcal{L}_{\text{CFM}}(\theta) \) and \( \mathcal{L}_{\text{FM}}(\theta) \) with respect to \( \theta \) are identical.
Building on optimal transport principles~\cite{DBLP:conf/aaai/OnkenF0R21}, the optimal-transport conditional flow matching (OT-CFM) refines CFM by enabling particularly simple gradient computations, making it more efficient for practical implementation. The OT-CFM loss function can be written as:
$$
\mathcal{L}_{\text{OT-CFM}}(\theta) = \mathbb{E}_{t,q(x_1),p_0(x_0)}||u_t(\phi _t(x)|x_1)-v_t(\phi _t(x);\theta)||^2
$$
The OT-flow is defined as \(\phi _t = (1-t)x_0 + tx_1\) which represents the flow from \( x_0 \) to \( x_1 \), where each data point \( x_1 \) is paired with a random sample \( x_0 \sim \mathcal{N}(0, I) \). The gradient vector field, whose expectation is the target function we aim to learn, is given by \( u_t(\phi_t(x_0) \mid x_1) = x_1 - x_0\). This vector field is linear, time-dependent, and just relies only on \( x_0 \) and \( x_1 \). These properties simplify and enhance training efficiency while improving generation speed and performance compared to DPMs.

In our proposed approach, we transform a random sample \(x_0\) from the standard Gaussian noise to \(x_1\), the target mel-spectrogram, under the condition of corresponding semantic tokens~\footnote{Semantic tokens are aligned to mel-spectrogram using repeated upsampling.} and the speaker embedding from a reference mel-spectrogram. Hence, the final loss can be described as:
$$
\mathcal{L}_{\text{FM}}(\theta)=\mathbb{E}_{t,q(x_1),p_0(x_0)}||(x_1-x_0)-v_t((1-t)x_0+tx_1,c;\theta)||^2
$$
where \(c\) represents an additional condition, which is the concatenation of semantic tokens and the speaker embeddings.
\subsection{Classifier-Free Guidance}
Classifier-Free Guidance (CFG)~\cite{DBLP:journals/corr/abs-2207-12598} replaces an explicit classifier with an implicit one, avoiding the need to compute the explicit classifier and the gradient. This approach has been demonstrated to improve the generation quality of diffusion probabilistic models. The final generation result can be guided by randomly dropping the conditioning signal during training and performing linear extrapolation between inference outputs with and without the condition \( c \). 
During generation, the vector field is modified as follows:
\[
v_{t,\text{CFG}} = (1+\alpha)\cdot v_t(\phi _t(x),c;\theta)  - \alpha \cdot v_t(\phi _t(x);\theta)
\]
where \(\alpha\) is the extrapolation coefficient of CFG.

\section{StreamFlow}
\begin{figure}[bth]
    \centering
    \includegraphics[width=\linewidth]{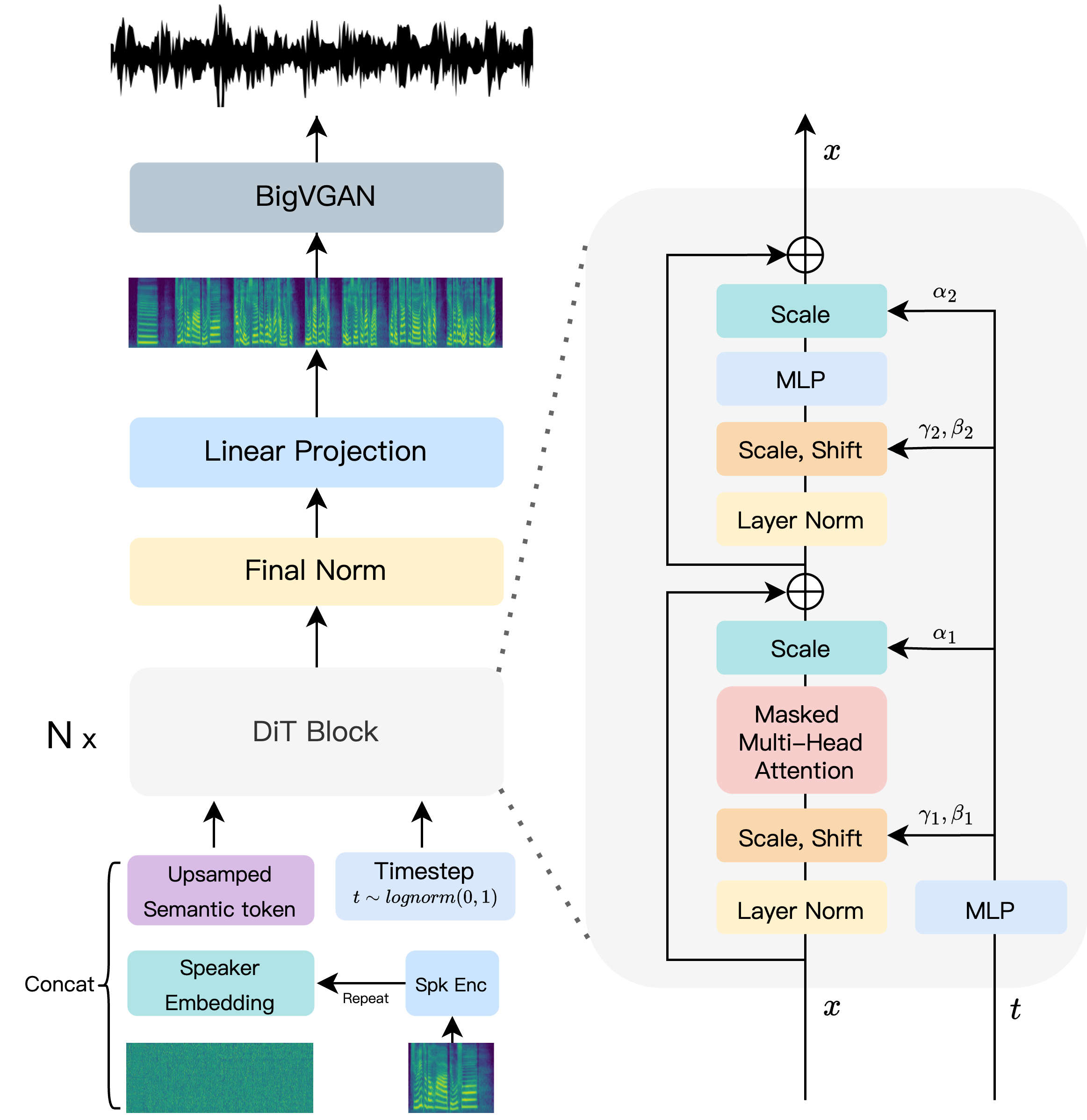}
    \caption{Overview architecture of our proposed StreamFlow.}
    \label{fig:Over}
\end{figure}

The overall architecture of StreamFlow is shown in Figure \ref{fig:Over}. We use the diffusion transformer (DiT) as the backbone, which consists of several DiT-blocks. To enhance stability and controllability during training, we apply zero-initialized adaptive LayerNorm (adaLN-zero) in the DiT-blocks. StreamFlow takes speech tokens and speaker embeddings as inputs, predicting the mel-spectrograms. We then use a BigVGAN~\cite{DBLP:conf/iclr/LeePGCY23} vocoder to upsample the predicted mel-spectrogram to 24kHz high-quality waveforms.

\subsection{Block-wise Guided Attention Masks}

\begin{figure}[h]
    \vspace{-7pt}
    \centering
    \includegraphics[width=0.9\linewidth]{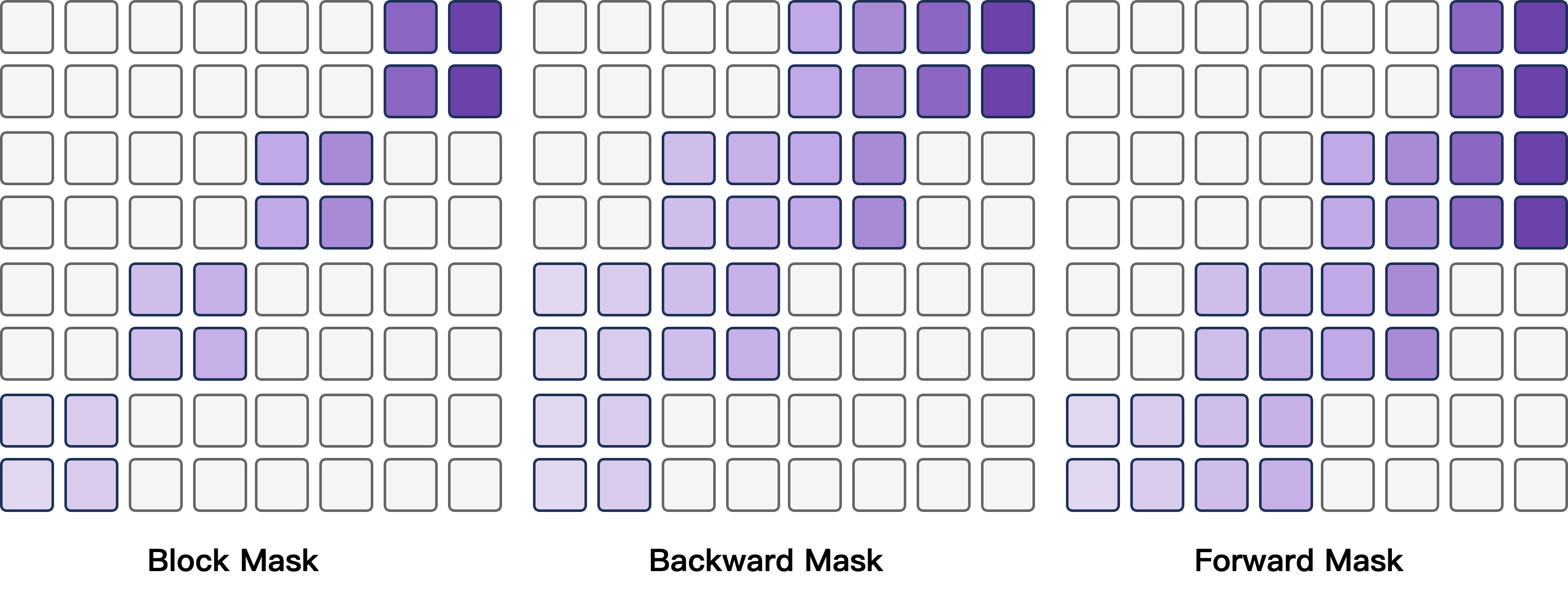}
    \vspace{-5pt}
    \caption{The details of the fundamental block-wise attention mask.}
    \label{fig:attn_mask}
    \vspace{-10pt}
\end{figure}

We consider that the generation from semantic tokens to waveforms can be formatted as a conversion process at localized. To capture this locality effectively, we group adjacent tokens into blocks, using these blocks as the fundamental units for implementing the block-wise guided attention mask. We assume the sequence length is \(n\) and the block size is \(b\) then the sequence can be divided into \(N_b =  \left \lfloor {n \over b} \right \rfloor \). For each position \(i \), we define the corresponding block as \(block(i) =  \left \lfloor {i \over b} \right \rfloor \). 
In this paper, we introduce three new fundamental block-wise attention masks, as shown in Figure \ref{fig:attn_mask}, enabling more flexible configurations for controlling future and past receptive fields across blocks.

\begin{itemize}
\item \textit{Block Mask }: Ensures that each block remains isolated, preventing interaction between them and preserving the original receptive field of the DiT model.
\item \textit{Backward Mask}: Allows each block to access the preceding block’s information, extending the DiT model’s receptive field forward by one block with each application.
\item \textit{Forward Mask}: Enables each block to access the subsequent block’s information, expanding the DiT model’s receptive field backward by one block with each application.
\end{itemize}

\begin{figure}[h]
    \centering
    \vspace{-10pt}
    \includegraphics[width=0.8\linewidth]{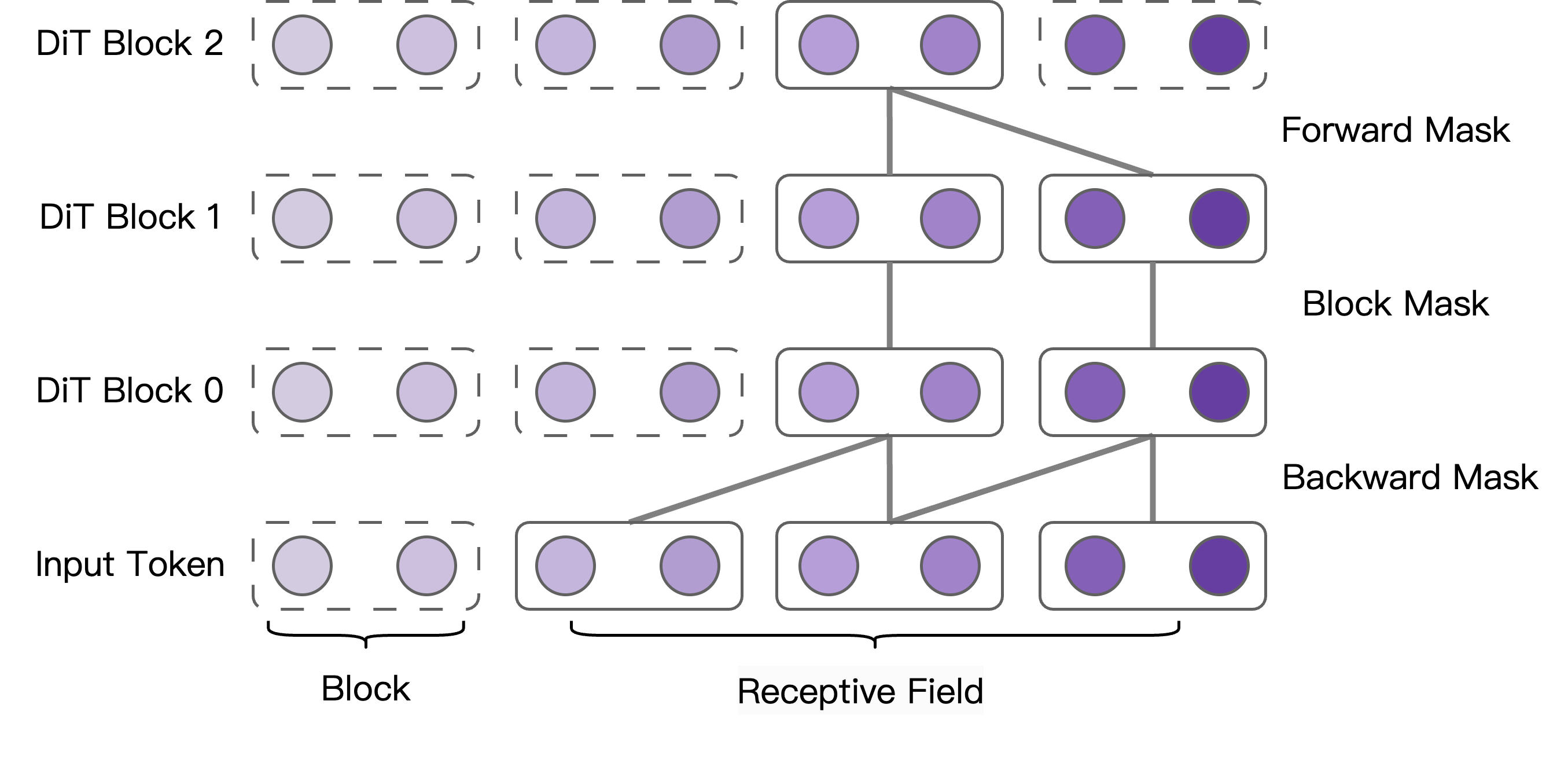}
    \vspace{-5pt}

    \caption{Receptive field of DiT model with 3 DiT-blocks. When all three masks are combined, the final receptive field spans all three blocks.}
    \label{fig:rf}
    \vspace{-15pt}
\end{figure}

Given a DiT model containing \(n\) DiT-blocks, if \(p\) blocks use the \textit{Backward Mask} and \(q\) blocks use the \textit{Forward Mask} while the rest use the \textit{Block Mask}, the overall receptive field of the model is \( (p+q + 1)\cdot b\) tokens~\footnote{\(q\) blocks for past and \(p\) blocks for future}. Figure \ref{fig:rf} is a simple case to demonstrate the receptive field with different masks.


\subsection{Streaming Waveform Generation}
\begin{figure}[h]
    \centering
    \vspace{-10pt}

    \includegraphics[width=0.75\linewidth]{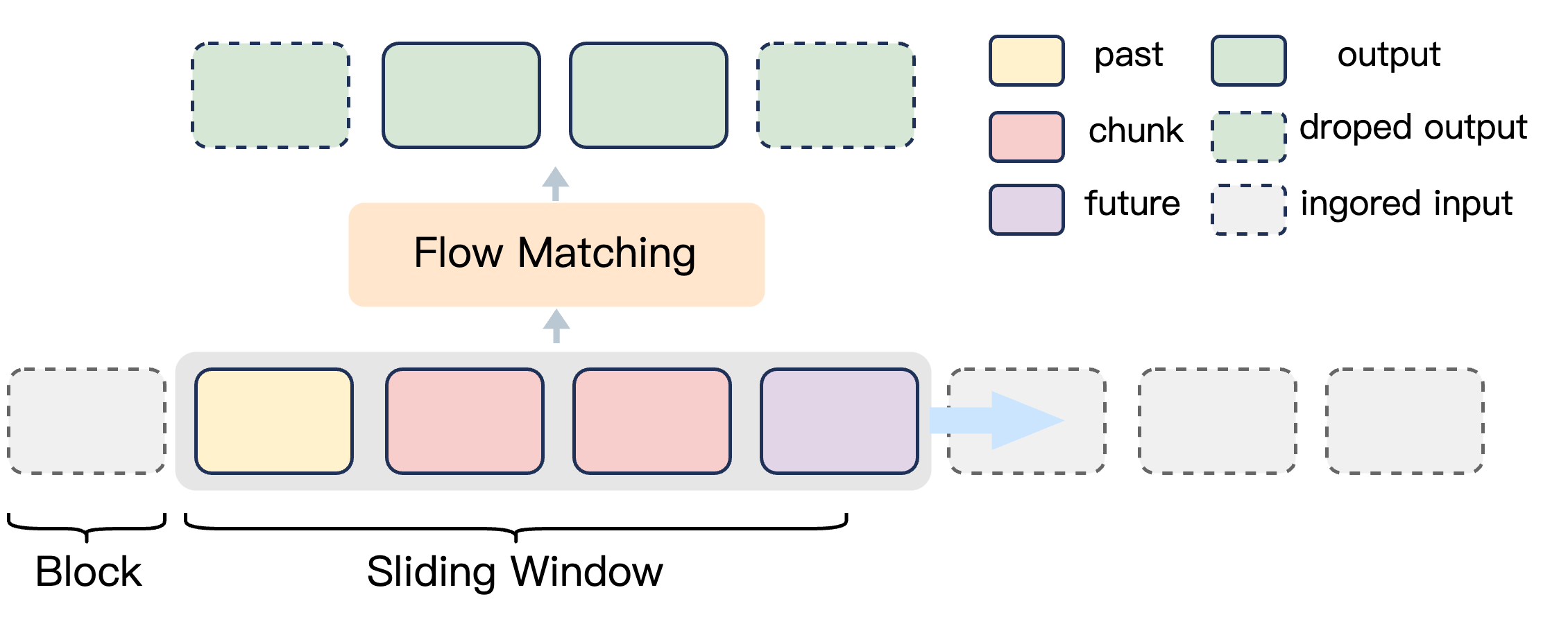}
    \vspace{-5pt}
    \caption{Streaming Inference process of StreamFlow.}
    \label{fig:infer}
    \vspace{-10pt}
\end{figure}

During inference, we implement a streaming inference process by adopting a chunk-by-chunk approach to generate mel-spectrograms using flow matching. As illustrated in Figure \ref{fig:infer}, the input sequence is partitioned into multiple blocks and subsequently grouped into chunks.  Each input chunk is carefully supplemented with preceding and succeeding contextual blocks that meet with the receptive field. These chunks are processed by the flow matching module in a sliding window manner, enabling the generation of mel-spectrograms while maintaining contextual consistency and smooth transitions across chunks. Additionally, since BigVGAN uses a convolutional architecture with a fixed receptive field, we apply a similar chunk-based processing strategy to BigVGAN, ensuring both real-time performance and high-quality waveform reconstruction.


\section{Experimental Setup}

\subsection{Training Setup}
For training, we use the Chinese and English subsets of Emilia~\cite{DBLP:journals/corr/abs-2407-05361}, which totals 100,000 hours of speech data.
For semantic tokens extraction, we employ 25Hz S3tokenizer\footnote{https://github.com/xingchensong/S3Tokenizer}, while 80-dimensional mel-spectrograms are extracted from 16kHz speech signals, with a frame size of 1024 and a hop size of 160. 

\subsection{Model Details}
We use the non-streaming flow matching (UNet + Transformer as the backbone) with HiFiGAN~\cite{DBLP:conf/nips/KongKB20} from CosyVoice\footnote{https://github.com/FunAudioLLM/CosyVoice} as a compare non-streaming model, named UNet-CV. For a fair comparison, we also implement both non-streaming and streaming approaches in CosyVoice based on the DiT backbone: (1) DiT-CV, which utilizes full attention for non-streaming generation; (2) DiT-CVS, which employs block-wise causal attention for streaming generation, similar to CosyVoice2.
Additionally, we implement two variants of StreamFlow with different receptive fields: StreamFlow-SR~(small receptive field), where each block has access to two previous blocks and one future block, with two backward masks at the 7th and 14th layers and one forward mask at the 1st layer; and StreamFlow-LR~(large receptive field), where each block has access to two previous blocks and two future blocks, with two backward masks at the 7th and 14th layers and two forward masks at the 1st and 22nd layers. 


In the experiment, a 22-layer DiT model is used as the backbone with a hidden dimension of 1024, 16 attention heads, and a dropout rate of 0.1, with approximately 330M parameters. The speaker encoder utilizes ECAPA-TDNN~\cite{DBLP:conf/interspeech/DawalatabadRGTD21} to extract utterance-level speaker embeddings. Logit-normal sampling is adopted instead of uniform sampling for timestep \(t\) to improve generation quality, as suggested in~\cite{DBLP:conf/icml/EsserKBEMSLLSBP24}. During CFG training, conditions are dropped at a rate of 0.3. For block-wise guided attention, the block size is set to 24 frames (0.24s). All DiT training is conducted on 16 NVIDIA A100 GPUs with a batch size of 12,800 frames, trained for 400,000 steps. For inference, exponential moving averaged (EMA) weights are used, and the Euler ODE solver is employed to compute the mel-spectrogram with 10 time steps, with a CFG strength of 0.5. The chunk size for all streaming models is set to 2 blocks.

\subsection{Evaluation Metric}

To evaluate the quality of generated speech, we use five objective metrics: (1) Short-Time Objective Intelligibility (STOI)~\cite{stoi} for speech intelligibility, (2) Perceptual Evaluation of Speech Quality (PESQ)~\cite{pesq} for speech quality, (3) Virtual Speech Quality Objective Listener (ViSQOL)~\cite{vosqol} for perceived audio quality, (4) UTMOS~\cite{utmos}, a widely used perceptual rating predictor, and (5) SECS, embedding cosine similarity pretrain-model\footnote{
https://github.com/BytedanceSpeech/seed-tts-eval
} for speaker similarity. The test set includes 100 randomly selected sentences from unseen speakers for generalizability. Additionally, we assess naturalness (NMOS) and speaker similarity (SMOS) Mean Opinion Score using 30 sentences in Chinese and English with 20 professional listeners.

\section{Experimental Results}

\subsection{Audio Quantity}
\begin{table}[h]
\vspace{-10pt}

\centering
\caption{Evaluation results for different flow matching models.}
\vspace{-5pt}

\label{obj}
\resizebox{\linewidth}{!}{

\begin{tabular}{c|c|cccccc}
\toprule
  Models        & Mode     & STOI$\uparrow$  & UTMOS$\uparrow$ & PESQ$\uparrow$  & ViSQOL$\uparrow$ & SECS$\uparrow$ \\
\midrule
UNet-CV    & \multirow{2}{*}{Non-Stream} & 0.827 & 3.671 & 1.361 & \textbf{4.102}  &  \textbf{ 0.743}  \\
DiT-CV      &  & \textbf{0.852} & \textbf{3.692} & \textbf{1.581} & 4.057  &   0.727  \\
\midrule
DiT-CVS      & \multirow{3}{*}{Stream} & 0.819 & 3.618 & 1.413 & 4.015  &   \textbf{0.717}  \\
StreamFlow-SR && \textbf{0.832} & \textbf{3.667} & 1.521 & \textbf{4.069}  &   0.709  \\
StreamFlow-LR && 0.829 & 3.638 & \textbf{1.531} & 4.054  &   0.721  \\
\bottomrule
\end{tabular}
}
\vspace{-15pt}

\end{table}

\subsubsection{Obeject Evaluation}
The objective results are shown in Table \ref{obj}, non-streaming models CosyVoice and DiT-CV exhibit similar audio quality, yet CosyVoice achieves higher speaker similarity due to its use of both in-context learning and speaker embedding while DiT-CV only uses speaker embedding. Meanwhile, DiT-based models outperform PESQ and STOI thanks to the superior audio reconstruction of BigVGAN over HiFiGAN. For streaming models, DiT-CVS performs slightly worse in audio quality because it only considers historical context; in contrast, StreamFlow-SR and StreamFlow-LR, which incorporate future information, yield better performance. Furthermore, StreamFlow-LR’s larger receptive field gives it a slight edge, underscoring the benefit of balancing historical and future information in streaming speech generation.

\begin{table}[h]
\vspace{-6pt}
\centering
\caption{Audio quality with different block sizes.}
\vspace{-5pt}

\label{obj-q}
\resizebox{0.8\linewidth}{!}{

\begin{tabular}{c|ccccccc}
\toprule
  Block Size             & STOI$\uparrow$  & UTMOS$\uparrow$ & PESQ$\uparrow$  & ViSQOL$\uparrow$ & SECS$\uparrow$ \\
\midrule
0.12s     & 0.806 & 3.571 & 1.361 & 3.940  &   \textbf{0.721}  \\
0.24s     & 0.832 & 3.667 & 1.521 & 4.069  &   0.701  \\
0.48s     & \textbf{0.848} & \textbf{3.698} & \textbf{1.573 }& \textbf{4.075}  &   0.713 \\
\bottomrule
\end{tabular}
}
\vspace{-5pt}

\end{table}
Further, we evaluate the performance of StreamFlow-SR with different block sizes. As shown in table \ref{obj-q}, increasing the block size leads to a noticeable improvement in audio quality, while speaker similarity remains largely unchanged. This is because increasing the block size also enlarges the model's receptive field.

\subsubsection{Subeject Evaluation}
As shown in table \ref{sbj}, the results of the subject evaluation align with the object evaluation. Specifically, CosyVoice outperforms DiT-based models in speaker similarity due to its use of in-context learning and speaker embedding. Moreover, increasing the receptive field in streaming models, as seen in StreamFlow-SR and StreamFlow-LR, leads to noticeable improvements in audio quality.

\begin{table}[bth]
\centering
\vspace{-8pt}
\caption{Subeject evaluation results for different flow matching models.}
\label{sbj}
\vspace{-5pt}

\resizebox{0.7\linewidth}{!}{

\begin{tabular}{c|c|cc}
\toprule
 Models      &Mode      & NMOS$\uparrow$  & SMOS$\uparrow$  \\
\midrule
 UNet-CV    &\multirow{2}{*}{Non-Stream} & \textbf{4.221 $\pm$ 0.07} &\textbf{ 4.371 $\pm$ 0.12}     \\
DiT-CV     &   & 4.213 $\pm$ 0.11 & 4.311 $\pm$ 0.10      \\
\midrule
DiT-CVS    & \multirow{3}{*}{Stream}  & 3.978 $\pm$ 0.10 & 4.193 $\pm$ 0.10     \\
StreamFlow-SR &   & 4.012 $\pm$ 0.09 &\textbf{ 4.221 $\pm$ 0.12}     \\
StreamFlow-LR &  & \textbf{4.153 $\pm$ 0.10 }& 4.152 $\pm$ 0.11     \\
\bottomrule
\end{tabular}
}
\vspace{-12pt}

\end{table}

\subsection{Inference Speed}

\begin{figure}[bth]
    \vspace{-10pt}

    \centering
    \includegraphics[width=0.95\linewidth]{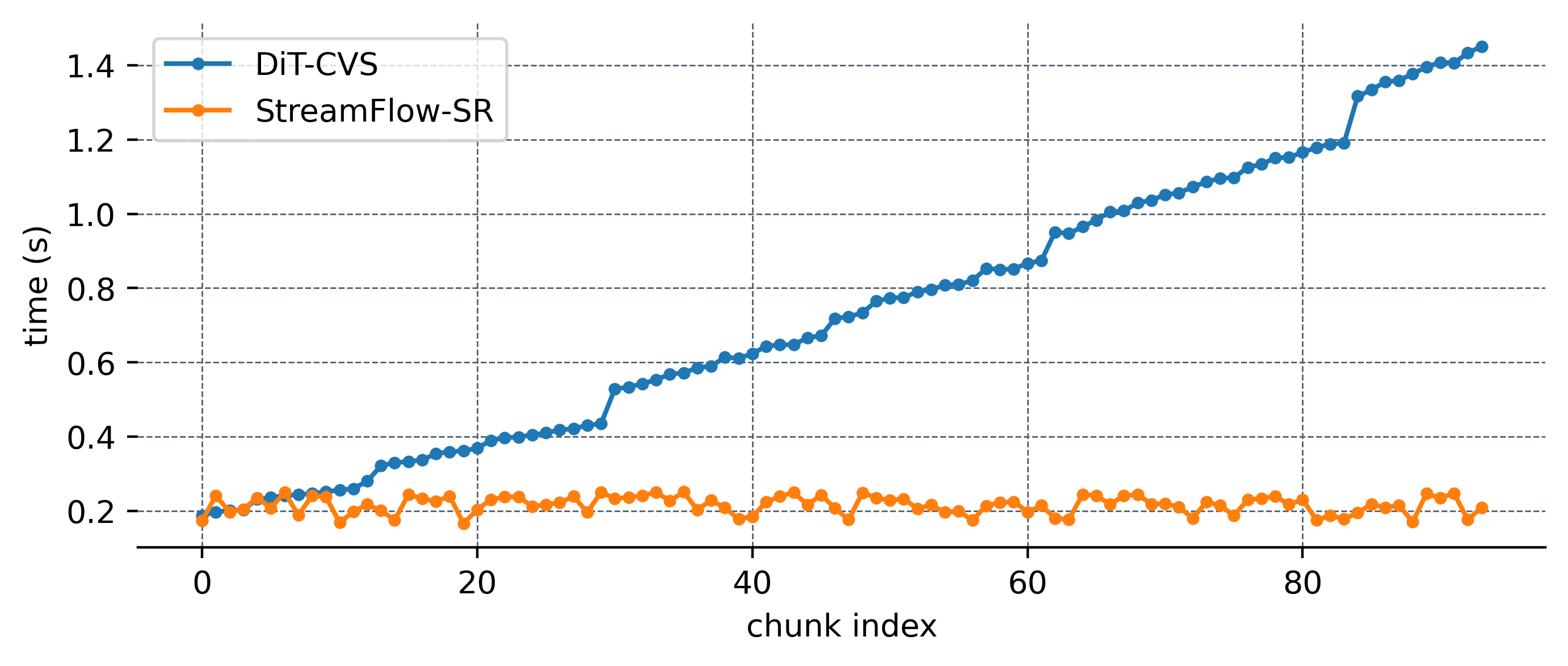}
    \vspace{-5pt}

    \caption{Chunk latency in long-form speech generation.}
    \label{fig:speed}
    \vspace{-10pt}

\end{figure}

To measure the inference speed, we calculate the latency of each chunk on the NVIDIA A100 GPU device.
As shown in Figure \ref{fig:speed}, we compare the inference speed of StreamFlow-SR and DiT-CVS in long-form speech generation scenarios. Both models achieve a first-packet latency of approximately 180ms. Since StreamFlow employs a sliding window mechanism, the computational cost per chunk remains constant, ensuring that inference time remains stable at around 180ms. In contrast, DiT-CVS incurs increasing computational overhead per chunk as historical information accumulates, leading to progressively longer inference times and reduced overall efficiency.

Moreover, both objective and subjective evaluations indicate that expanding the receptive field improves model performance. However, in practical applications, when integrating with Codec-LM, a larger receptive field, paralytically future receptive field, necessitates generating more tokens, thereby increasing computational costs and further extending the system’s first-packet latency. Therefore, a balance must be struck between synthesis quality and inference efficiency to maintain high-generation quality and real-time performance in low-latency scenarios.
\section{Conclusions}

In this paper, we propose StreamFlow, a streaming speech token decoding framework based on DiT-driven flow matching, addressing the challenges of real-time speech generation while maintaining high quality. 
We design a local block-wise receptive field method to achieve efficient streaming inference, particularly for long-speech scenarios.
Experimental results demonstrate that our proposed StreamFlow achieves comparable performance to non-streaming methods in both objective and subjective evaluations while significantly reducing inference latency during long-sequence generation. Moreover, our model achieves a first-packet latency of only 180ms, meeting the real-time requirements of interactive speech applications.
\bibliographystyle{IEEEtran}
\bibliography{mybib}

\begin{thebibliography}{10}
\providecommand{\url}[1]{#1}
\csname url@samestyle\endcsname
\providecommand{\newblock}{\relax}
\providecommand{\bibinfo}[2]{#2}
\providecommand{\BIBentrySTDinterwordspacing}{\spaceskip=0pt\relax}
\providecommand{\BIBentryALTinterwordstretchfactor}{4}
\providecommand{\BIBentryALTinterwordspacing}{\spaceskip=\fontdimen2\font plus
\BIBentryALTinterwordstretchfactor\fontdimen3\font minus \fontdimen4\font\relax}
\providecommand{\BIBforeignlanguage}[2]{{%
\expandafter\ifx\csname l@#1\endcsname\relax
\typeout{** WARNING: IEEEtran.bst: No hyphenation pattern has been}%
\typeout{** loaded for the language `#1'. Using the pattern for}%
\typeout{** the default language instead.}%
\else
\language=\csname l@#1\endcsname
\fi
#2}}
\providecommand{\BIBdecl}{\relax}
\BIBdecl

\bibitem{DBLP:journals/corr/abs-2410-21276}
OpenAI, ``{GPT-4o} system card,'' \emph{CoRR}, vol. abs/2410.21276, 2024.

\bibitem{DBLP:journals/corr/abs-2410-00037}
A.~D{\'{e}}fossez, L.~Mazar{\'{e}}, M.~Orsini, A.~Royer, P.~P{\'{e}}rez, H.~J{\'{e}}gou, E.~Grave, and N.~Zeghidour, ``Moshi: a speech-text foundation model for real-time dialogue,'' \emph{CoRR}, vol. abs/2410.00037, 2024.

\bibitem{DBLP:conf/iclr/0006H0QZZL21}
Y.~Ren, C.~Hu, X.~Tan, T.~Qin, S.~Zhao, Z.~Zhao, and T.~Liu, ``Fastspeech 2: Fast and high-quality end-to-end text to speech,'' in \emph{Proc. {ICLR}}, 2021.

\bibitem{DBLP:conf/icml/KimKS21}
J.~Kim, J.~Kong, and J.~Son, ``Conditional variational autoencoder with adversarial learning for end-to-end text-to-speech,'' in \emph{Proc. {ICML}}, 2021, pp. 5530--5540.

\bibitem{DBLP:journals/corr/abs-2410-23815}
D.~Guo, J.~Yao, X.~Zhu, K.~Xia, Z.~Guo, Z.~Zhang, Y.~Wang, J.~Liu, and L.~Xie, ``The {NPU-HWC} system for the {ISCSLP} 2024 inspirational and convincing audio generation challenge,'' 2024.

\bibitem{DBLP:journals/corr/abs-2305-07243}
J.~Betker, ``Better speech synthesis through scaling,'' \emph{CoRR}, vol. abs/2305.07243, 2023.

\bibitem{DBLP:journals/corr/abs-2402-08093}
M.~Lajszczak, G.~C{\'{a}}mbara, Y.~Li, F.~Beyhan, A.~van Korlaar, F.~Yang, A.~Joly, {\'{A}}.~Mart{\'{\i}}n{-}Cortinas, A.~Abbas, A.~Michalski, A.~Moinet, S.~Karlapati, E.~Muszynska, H.~Guo, B.~Putrycz, S.~L. Gambino, K.~Yoo, E.~Sokolova, and T.~Drugman, ``{BASE} {TTS:} lessons from building a billion-parameter text-to-speech model on 100k hours of data,'' \emph{arXiv preprint arXiv:2402.08093}, 2024.

\bibitem{DBLP:journals/corr/abs-2411-13577}
S.~Ji, Y.~Chen, M.~Fang, J.~Zuo, J.~Lu, H.~Wang, Z.~Jiang, L.~Zhou, S.~Liu, X.~Cheng, X.~Yang, Z.~Wang, Q.~Yang, J.~Li, Y.~Jiang, J.~He, Y.~Chu, J.~Xu, and Z.~Zhao, ``Wavchat: {A} survey of spoken dialogue models,'' \emph{CoRR}, vol. abs/2411.13577, 2024.

\bibitem{cosy}
Z.~Du, Q.~Chen, S.~Zhang, K.~Hu, H.~Lu, Y.~Yang, H.~Hu, S.~Zheng, Y.~Gu, Z.~Ma, Z.~Gao, and Z.~Yan, ``Cosyvoice: {A} scalable multilingual zero-shot text-to-speech synthesizer based on supervised semantic tokens,'' \emph{CoRR}, vol. abs/2407.05407, 2024.

\bibitem{DBLP:journals/corr/abs-2412-02612}
A.~Zeng, Z.~Du, M.~Liu, K.~Wang, S.~Jiang, L.~Zhao, Y.~Dong, and J.~Tang, ``Glm-4-voice: Towards intelligent and human-like end-to-end spoken chatbot,'' \emph{CoRR}, vol. abs/2412.02612, 2024.

\bibitem{DBLP:conf/iclr/LipmanCBNL23}
Y.~Lipman, R.~T.~Q. Chen, H.~Ben{-}Hamu, M.~Nickel, and M.~Le, ``Flow matching for generative modeling,'' in \emph{Proc. {ICLR}}, 2023.

\bibitem{cosy2}
Z.~Du, Y.~Wang, Q.~Chen, X.~Shi, X.~Lv, T.~Zhao, Z.~Gao, Y.~Yang, C.~Gao, H.~Wang, F.~Yu, H.~Liu, Z.~Sheng, Y.~Gu, C.~Deng, W.~Wang, S.~Zhang, Z.~Yan, and J.~Zhou, ``Cosyvoice 2: Scalable streaming speech synthesis with large language models,'' \emph{CoRR}, vol. abs/2412.10117, 2024.

\bibitem{DBLP:conf/iccv/PeeblesX23}
W.~Peebles and S.~Xie, ``Scalable diffusion models with transformers,'' in \emph{Proc. ICCV}, 2023.

\bibitem{DBLP:conf/nips/ChenRBD18}
T.~Q. Chen, Y.~Rubanova, J.~Bettencourt, and D.~Duvenaud, ``Neural ordinary differential equations,'' in \emph{Proc. NeurIPS}, S.~Bengio, H.~M. Wallach, H.~Larochelle, K.~Grauman, N.~Cesa{-}Bianchi, and R.~Garnett, Eds., 2018.

\bibitem{DBLP:conf/aaai/OnkenF0R21}
D.~Onken, S.~W. Fung, X.~Li, and L.~Ruthotto, ``Ot-flow: Fast and accurate continuous normalizing flows via optimal transport,'' in \emph{Proc. AAAI}, 2021, pp. 9223--9232.

\bibitem{DBLP:journals/corr/abs-2207-12598}
J.~Ho and T.~Salimans, ``Classifier-free diffusion guidance,'' \emph{CoRR}, vol. abs/2207.12598, 2022.

\bibitem{DBLP:conf/iclr/LeePGCY23}
S.~Lee, W.~Ping, B.~Ginsburg, B.~Catanzaro, and S.~Yoon, ``Bigvgan: {A} universal neural vocoder with large-scale training,'' in \emph{Proc. {ICLR}}, 2023.

\bibitem{DBLP:journals/corr/abs-2407-05361}
H.~He, Z.~Shang, C.~Wang, X.~Li, Y.~Gu, H.~Hua, L.~Liu, C.~Yang, J.~Li, P.~Shi, Y.~Wang, K.~Chen, P.~Zhang, and Z.~Wu, ``Emilia: An extensive, multilingual, and diverse speech dataset for large-scale speech generation,'' \emph{CoRR}, vol. abs/2407.05361, 2024.

\bibitem{DBLP:conf/nips/KongKB20}
J.~Kong, J.~Kim, and J.~Bae, ``Hifi-gan: Generative adversarial networks for efficient and high fidelity speech synthesis,'' in \emph{Proc. NeurIPS}, H.~Larochelle, M.~Ranzato, R.~Hadsell, M.~Balcan, and H.~Lin, Eds., 2020.

\bibitem{DBLP:conf/interspeech/DawalatabadRGTD21}
N.~Dawalatabad, M.~Ravanelli, F.~Grondin, J.~Thienpondt, B.~Desplanques, and H.~Na, ``{ECAPA-TDNN} embeddings for speaker diarization,'' in \emph{Proc. Interspeech}, H.~Hermansky, H.~Cernock{\'{y}}, L.~Burget, L.~Lamel, O.~Scharenborg, and P.~Motl{\'{\i}}cek, Eds., 2021.

\bibitem{DBLP:conf/icml/EsserKBEMSLLSBP24}
P.~Esser, S.~Kulal, A.~Blattmann, R.~Entezari, J.~M{\"{u}}ller, H.~Saini, Y.~Levi, D.~Lorenz, A.~Sauer, F.~Boesel, D.~Podell, T.~Dockhorn, Z.~English, and R.~Rombach, ``Scaling rectified flow transformers for high-resolution image synthesis,'' in \emph{Proc. {ICML}}, 2024.

\bibitem{stoi}
C.~H. Taal, R.~C. Hendriks, R.~Heusdens, and J.~Jensen, ``An algorithm for intelligibility prediction of time-frequency weighted noisy speech,'' \emph{{IEEE} Trans. Speech Audio Process.}, pp. 2125--2136, 2011.

\bibitem{pesq}
A.~W. Rix, J.~G. Beerends, M.~P. Hollier, and A.~P. Hekstra, ``Perceptual evaluation of speech quality (pesq)-a new method for speech quality assessment of telephone networks and codecs,'' in \emph{Proc. {ICASSP}}, 2001.

\bibitem{vosqol}
A.~Hines, J.~Skoglund, A.~C. Kokaram, and N.~Harte, ``Visqol: an objective speech quality model,'' \emph{{EURASIP} J. Audio Speech Music. Process.}, vol. 2015, p.~13, 2015.

\bibitem{utmos}
T.~Saeki, D.~Xin, W.~Nakata, T.~Koriyama, S.~Takamichi, and H.~Saruwatari, ``{UTMOS:} utokyo-sarulab system for voicemos challenge 2022,'' in \emph{Proc. Interspeech}, 2022.

\end{thebibliography}

\end{document}